\documentclass[pop,fleqn]{w-art}
\usepackage{times}
\usepackage{w-thm}
\theoremstyle{plain}

\theoremstyle{definition}

\usepackage{graphicx}
\usepackage{amsmath,amssymb,graphicx,amsthm}
\begin{document}

\numberwithin{equation}{section}
\newcommand{\der}{\partial}
\newcommand{\e}{\mathrm{e}}
\newcommand{\ii}{\mathrm{i}}
\newcommand{\dd}{\mathrm{d}}
\newcommand{\vf}{\varphi}
\newcommand{\eps}{\varepsilon}
\newcommand{\RR}{\mathbb R}
\newcommand{\NN}{\mathbb N}
\newcommand{\mM}{\mathcal{M}}
\newcommand{\mL}{\mathcal{L}}
\newcommand{\tend}{\rightarrow}
\newcommand{\then}{\Rightarrow}
\newcommand{\Tr}{\mathrm{Tr}}
\newcommand{\sla}{\!\!\!\!\slash}
\newcommand{\Ree}{\Re\mathrm{e}\,}
\newcommand{\Imm}{\Im\mathrm{m}\,}
\newcommand{\ve}[2]{e^{#1}_{\phantom{#1}{#2}}}
\newcommand{\veps}[2]{{\varepsilon}^{#1}_{\phantom{#1}{#2}}}
\newcommand{\sigmahat}[2]{{\sigma}^{\hat #1}_{\phantom{#1}{\hat #2}}}
\newcommand{\sigmatilde}[2]{{\sigma}^{\tilde #1}_{\phantom{#1}{\tilde #2}}}
\newcommand{\vE}[2]{E^{#1}_{\phantom{#1}{#2}}}
\newcommand{\Sigmahat}[2]{{\Sigma}^{\hat #1}_{\phantom{#1}\hat {#2}}}
\newcommand{\Sigmatilde}[2]{{\Sigma}^{\tilde #1}_{\phantom{#1}{\tilde #2}}}
\newcommand{\sth}{\sigma_{\hat 3}}




\title[BPS solutions in AdS/CFT]{BPS solutions in AdS/CFT}

\address[\inst{1}]{High Energy Section,
The Abdus Salam International Centre for Theoretical Physics,
Strada Costiera 11, 34014 Trieste, Italy}
\address[\inst {2}]{Istituto Nazionale di Fisica Nucleare, sez. di Trieste}
\address[\inst{3}]{Scuola Internazionale Superiore di Studi
Avanzati,
Via Beirut 2-4, 34014 Trieste, Italy}
\address[\inst{4}]{Univerza v Novi Gorici,
Vipavska 13, P.P. 301,
Ro\v{z}na dolina,
SI-5000 Nova Gorica, Slovenia} 
\author[E. Gava]{Edi Gava\inst{1,2}}
\author[G. Milanesi]{Giuseppe Milanesi\inst{2,3}%
  \footnote{Corresponding author\quad E-mail:~\textsf{milanesi@sissa.it},
}}

\author[K.S. Narain]{Kumar S. Narain\inst{1}}
\author[M. O'Loughlin]{Martin O'Loughlin\inst{1,4}}

\begin{abstract}
We study a class of exact supersymmetric solutions of type IIB Supergravity. 
They have an $SO(4)\times SU(2) \times U(1)$ isometry and preserve generically 
4 of the 32 supersymmetries of the theory. Asymptotically $AdS_5 \times S^5$ 
solutions in this class are dual to 1/8 BPS chiral operators which
preserve the same symmetries in the $\mathcal N=4$ SYM theory. They are parametrized by a 
set of four functions that satisfy certain differential equations. 
We analyze the solutions to these equations in a large radius asymptotic expansion:
they carry charges with respect to two $U(1)$  KK gauge fields
and their mass saturates the expected BPS bound. We also show how the same formalism is suitable for the description of excitations of the $AdS_5\times Y^{p,q}$ geometries.
\end{abstract}
\maketitle   

\section{Introduction}

 One of the first steps in understanding the $AdS / CFT$ 
correspondence is to set up a precise dictionary  between the states 
of the  theories on the two sides of the correspondence. 
It is well known that the parameters of the $\mathcal N = 4$ $SU(N)$ SYM, 
namely $\lambda \equiv g_{YM}^2 N$ and $N$ should be identified 
with the parameters of type IIB String Theory on 
$AdS_5\times S^5$, namely $L_{AdS},\ell_s,g_s$, as $ L_{AdS}^4 = 4 \pi\lambda \ell_s^4 $, $g_s = \lambda/N$.

We may consider solutions to the supergravity equations 
of motion which are asymptotically $AdS_5 \times S^5$ as good 
candidates for dual of states in the CFT,  provided that 
$N\gg \lambda \gg 1$ and  any dimension four  curvature 
invariant of the solutions, $ \mathcal R_4$, satisfies
\begin{equation}
	\mathcal R_4  L_{AdS}^4 \ll \lambda \ll N\,.
\end{equation}
Here and in \cite{Gava:2006pu} we address the problem of finding BPS 
supergravity solutions which are dual to chiral primary operators of the form
\begin{equation}\label{gaugestates}
	\sum c_i \Tr \big( Z_1^q\big) \Tr  \big(Z_2^q\big)\Tr \big( Z_3^{r_i} \big)\,,
\end{equation}
where $Z_1,Z_2$ and $Z_3$ are the three complex scalars of the 
$\mathcal N =4$ CFT and the coefficients $c_i$ are chosen to give a (anti)-symmetrization and trace structure such that the $s$-wave mode on $S^3$ of these operators are BPS.
The corresponding states saturate the BPS bound:
\begin{equation}
	\Delta = 2q+r\,,
\end{equation}
where $\Delta$ is the conformal dimension of the operator.
We will also assume that the index structure is such that the states preserve an 
$SU(2)_L \times U(1)_R$ subgroup of the $SO(6)$ $R$-symmetry group.
The full amount of bosonic symmetry preserved by these states is:
\begin{equation}
	\RR_{BPS}\times \big(SU(2)_L \times 
U(1)_R\big)_{R- \textrm{charge}}\times SO(4)_{KK}\,.
\end{equation}
The first $\RR$ corresponds to the transformations generated by $D^\prime \equiv D - 2 J^3_R - J\,,$
where $J=J_3$ generates phase shifts on $Z_3$, $J_3^R$ generates simultaneous equal phase shifts on $Z_1, Z_3$, $D$ is the dilatation operator
and the last $SO(4)$ factor represents the fact that 
we are considering $s$-wave modes on $S^3$ in the reduction
of SYM theory on $\RR\times S^3$ \cite{Witten:1998qj}.
Consequently, we start from an 
Ansatz for the metric and the self-dual RR 5-form 
which preserves this amount of symmetry. 
We also require that the background possesses the required 
amount of supersymmetry by demanding that it possesses a Killing spinor. 

\section{Supergravity analysis}
We look for geometries dual to the states in \eqref{gaugestates}. They will thus be invariant under $SU(2)_L \times SO(4)_{KK}$ as defined in the previous section and invariant but charged under the remaining $U(1)_R$. The extra non-compact time-like 
symmetry ($\RR_{BPS}$ of the previous section) is associated to invariance under 
the transformations generated by $D^\prime$ in the 
gauge theory and will emerge as a direct consequence of supersymmetry.

The most generic Ansatz consistent with these symmetries is given by
\begin{equation}
	\dd s^2 = g_{\mu\nu} \dd x^\mu \dd x^\nu + \rho_1^2 \big(\hat \sigma_1^2 + \hat\sigma_2^2\big) +\rho_3^2(\hat\sigma_3 - A_\mu \dd x^\mu)^2 + \tilde \rho^2 \dd \tilde \Omega_3^2\,.
\end{equation}
where $\rho_1$, $\rho_3$, $\tilde{\rho}$, $A_{\mu}$ and $g_{\mu \nu}$ are functions of the four
coordinates $x^{\mu}$.
The space is a fibration of a squashed 3-sphere (on which the $SU(2)$ left-invariant 1-forms $\sigma^{\hat a}$ are defined) and a round 3-sphere over a four dimensional manifold.

Since we are looking for the geometric dual to operators which involve only scalar fields 
in the gauge theory, the only possible non-zero Ramond-Ramond field strength is the 
five form $F_{(5)}$. The most generic Ansatz for the five form which is 
invariant under the given symmetries is:
\begin{equation}
 	 F_{(5)} = \mathcal F_5 + \star \mathcal F_5
\end{equation}
with
\begin{equation}
 	\frac 12 \mathcal F_5= \left[\tilde G_{\mu\nu} \dd x^\mu\wedge \dd x^\nu + \tilde V_\mu \dd x^\mu\wedge (\hat\sigma_3 - A_\mu\dd x^\mu )+ \tilde g \hat\sigma_1\wedge \hat\sigma_2 \right]\wedge \tilde\rho^3\dd \tilde \Omega_3
\end{equation}

Supersymmetry is guaranteed by the existence of a supersymmetry parameter $\psi$ such that the 
gravitino variation it generates vanishes:
\begin{equation}
	\delta \chi_M = \nabla_M \psi + \frac{\ii}{480}F_{M_1 M_2 M_3 M_4 M_5} \Gamma^{M_1 M_2 M_3 M_4 M_5}\Gamma_M \psi = 0 \,.
\end{equation}
The vector $K=\bar\psi\Gamma^\mu\der_\mu\psi$ is Killing and timelike and it generates the extra non-compact timelike $U(1)$. We may thus choose the time coordinate $t$ such that $\der_t = K$.
The Bianchi  identity $\dd F_{(5)}=0$ and the existence of the spinor $\psi$ are sufficient for our supergravity 
background to satisfy the full equations of motion of type IIB Supergravity and to express the complete solution in the following form \cite{Gava:2006pu}:
\begin{multline}
\dd s^2 = - h^{-2} (\dd t + V_i \dd x^i)^2 + h^2 \frac{\rho_1^2}{\rho_3^2}(T^2\delta_{ij}\dd x^i\dd x^j + \dd y^2)+\tilde\rho^2\dd\tilde\Omega_3^2+ \\
+ \rho_1^2 \big(\hat\sigma_1^2 + \hat\sigma_2^2\big) +\rho_3^2(\hat\sigma_3 - A_t \dd t - A_i \dd x^i)^2
\end{multline}
where the coordinate $y$ is the product of two radii
\begin{equation}
y=\rho_1 \tilde \rho > 0\,.
\end{equation}
All the entries of the metric depend only on $(x^1,x^2,y)$ and can be expressed in terms of four independent functions $m,n,p,T$ as follows:
\begin{gather*}	
\begin{array}{lll}
 \rho_1^4 =  \frac{m p+ n^2}{m} y^4& \rho_3^4 = \frac{p^2}{m(mp+n^2)}&\tilde\rho^4 = \frac{m}{mp+n^2}\\ 
  h^{4} = \frac{ m p^2}{mp+n^2}& A_t = \frac{n-p}{p} & A_i = A_t V_i -\frac12 \epsilon_{ij}\der_j \ln T\\
\end{array}\\
\dd V = -y \star_3 [ \dd n + (n D +2y m (n-p) + 2n/y)\dd y]\label{firstdV}
\end{gather*}
where $\star_3$ indicates the Hodge dual in the three dimensional diagonal metric
$\dd s_3^2 = T^2\delta_{ij}\dd x^i\dd x^j + \dd y^2$.
Similarly, one can completely express the 5-form in terms of $m,n,p,T$ \cite{Gava:2006pu}. 
The complete analysis in \cite{Gava:2006pu} show that these solutions preserve generically 4 of the 32 supersymmetries of type IIB Supergravity.

The Bianchi identities and integrability conditions give rise to a set of nonlinear coupled elliptic differential equations
\begin{equation}\label{mnpT}
\begin{split}
&y^3 (\der_1^2+\der_2^2) n + \der_y \left(y^3 T^2 \der_y n\right) + y^2\der_y\big[ T^2 \big(y D n +2y^2m(n-p)\big)\big]+4 y^2 D T^2 n=0\\
&y^3 (\der_1^2+\der_2^2) m + \der_y \left(y^3 T^2 \der_y m\right) + \der_y \left( y^3 T^2 2m D\right)=0\\
&y^3 (\der_1^2+\der_2^2) p + \der_y \left(y^3 T^2 \der_y p\right) + \der_y \big[ y^3 T^2 4n y(n-p)\big]=0\\
&\der_y \ln T = D\qquad  D\equiv 2y(m+n-1/y^2) \,.
\end{split}
\end{equation}

\section{Solution to the Supergravity equations}
A solution to the equations presented in the previous section is determined by a set of boundary conditions at 
infinity (large values of $y,x^i$) and on the plane $y=0$; they should 
be chosen in such a way as to give an asymptotically to  $AdS_5\times S^5$ non-singular geometry.
Due to the non-linearity of the equations the relationship between  
boundary conditions and non-singular solutions with $AdS_5\times S^5$ asymptotics 
is difficult to control.

The solutions described in \cite{bubbling}, are a subset of ours specified by the additional constraints,
\begin{equation}
	n=p=\frac 1{y^2} -m = \frac{1/2-z}{y^2}\quad T=1\,.
\end{equation}
In this case we have
\begin{equation}
	D=0\quad\rho_1=\rho_3=\rho\quad A_t = 0\quad T=1
\end{equation}
and the three second order equations collapse to one single linear equation. As this 
equation is linear it has been possible to completely identify the boundary conditions at $y=0$ 
and at infinity that give rise to regular asymptotically $AdS_5\times S^5$ 
geometries\cite{bubbling,Milanesi:2005tp}.

\section{Asymptotics and charge}
In this section we discuss asymptotic solutions to the differential equations in \eqref{mnpT} for large values of $(x^1,x^2,y)$ which give $AdS_5 \times S^5$ asymptotics. 

The first corrections to the $AdS_5\times S^5$ geometry capture the global $U(1)$ charges 
under two  out of the three  $U(1)$ Cartan gauge fields arising from the KK reduction of  IIB supergravity on $S^5$ to five dimensional maximal gauged supergravity.

It is not hard to see that the following expressions for our functions
\begin{equation}\label{asympexp}
	m \sim \frac 1{y^2} - \frac{1}{L^4 R^4}\,,\quad n \sim \frac{1}{L^4 R^4}\,,\quad p \sim \frac{1}{L^4 R^4}\,,\quad T\sim 1
\end{equation}
satisfy the equations at leading order for large $R$, with $(L^2 R,\theta,\phi)$ polar coordinates in the $(x_1,x_2,y)$ space and give the leading asymptotics
\begin{equation}
	\dd s^2 = L^2 \left(- \tilde R^2  \dd t^2 + \frac{\dd \tilde R^2}{\tilde R^2} +  {\tilde R^2}\dd \tilde\Omega_3^2+\dd \theta^2  + \cos^2\theta \dd\tilde \phi^2 + \sin^2\theta\dd \hat\Omega_3^2\right)
\end{equation}
with $\tilde \phi = \phi-t $, which is $AdS_5\times S^5$ in Poincare coordinates.

We will now solve the differential equations to the next two orders in the asymptotic expansion. For the sake of simplicity we will assume that $\der_\phi$ is also a Killing vector of our solutions. Despite this simplifying assumption, in general the solutions will still be charged under the corresponding KK gauge field. 

We assume the following expansion for our functions:
\begin{equation}\label{asympexpsub}
\begin{split}
	&m \sim \frac 1{y^2} - \frac{1}{L^4R^4}+\frac{m_2(\theta)}{R^6}+\frac{m_3(\theta)}{R^8}\\
 &n \sim \frac{1}{L^4R^4}+\frac{n_2(\theta)}{R^6}+\frac{n_3(\theta)}{R^8}\\
 &p \sim \frac{1}{L^4R^4}+\frac{p_2(\theta)}{R^6}+\frac{p_3(\theta)}{R^8}\\
	&T\sim 1 + \frac{t_2(\theta)}{R^4}\,.
\end{split}
\end{equation}
At each order, the solutions to the equations \eqref{mnpT} will depend generically on seven integration constants. For the first subleading order, regularity of the asymptotic solution and coordinate redefinitions reduce these to two, $J$ and $Q$.

The global $U(1)$ charges 
under the Cartan gauge fields arising in the Kaluza Klein reduction of IIB supergravity over $S^5$ can be read off from the off diagonal components of the metric 
\begin{equation}
	g_{t \omega}\propto g_{\omega\omega} \frac{1}{R^2} 
\end{equation}
where $\omega$ is a suitable angle in the $S^5$ and the constant of proportionality is the global charge. The constants $J$ and $Q$ turn out to be the global charges under the $U(1)$ gauge fields dual to $J^3_R$ and $J$ of the gauge theory side. From the $g_{tt}$ we can read off the mass $M$ of our excitations which correctly satisfy the expected BPS bound
\begin{equation}
 		M = \frac{\pi L_{AdS}^2}{4 G_5}(|J|+2| Q|)\,.
\end{equation}

\section{Sasaki-Einstein manifolds}
In the previous section we assumed that $\der_\phi$ is a Killing vector of our metric. This lifts the internal symmetry to $SU(2)\times U(1)\times U(1)$. We can further restrict our solutions requiring that they preserve 8 instead of only 4 supersymmetries. This results in a set of constraints on the functions $m,n,p,T$ and give rise to $AdS_5\times Y^{p,q}$ geometries, where $Y^{p,q}$ are the Sasaki-Einstein manifolds constructed in \cite{Gauntlett:2004yd}. 
Type IIB String Theory on $AdS_5\times Y^{p,q}$ is conjectured to be dual to an $\mathcal N=1$ Quiver gauge theory specified by the 2 integers $p,q$. It is clear that, within our formalism, we can describe asymptotically $AdS_5\times Y^{p,q}$ solutions which are dual to operators in the corresponding quiver theory which preserve one half of its supersymmetry. 

We can show an explicit example for the case of $T^{1,1} \equiv Y^{1,0}$ \cite{Herzog:2002ih}. The corresponding gauge theory has an $SU(N)\times SU(N)$ gauge group, and two $SU(2)$ doublets $A,B$ of chiral superfields transforming in the $(N,\bar N)$ and $(\bar N, N)$ respectively of the gauge group. The $SU(2)$ is a flavour symmetry. In the case of $T^{1,1}$ the $SU(2)\times U(1)\times U(1)$ symmetry is lifted to $SU(2)\times SU(2)\times U(1)$ as a reflection of this flavour symmetry. Assuming the following leading behaviour for the functions $m,n,p,T$ 
\begin{gather*}
	m \sim \frac 3 2\frac 1{y^2} \quad n \sim \frac 23\frac{L^4}{y^4} \quad p \sim \frac 49 \frac{L^4}{y^4}\quad
	T\sim \frac {y}{L^2} \sqrt 6 \sin(\theta) \left(\tan\frac \theta 2\right)^{-1/6}
\end{gather*}
with $y^2 = \frac 2 3 L^2 \tilde \rho^2$ and $x_1^2+x_2^2 = L^4 \left(\tan\frac \theta 2\right)^{-1/3}$ and a proper parametrisation of the left invariant one forms, one obtains the following expression for the leading order metric
\begin{multline}
\dd s^2 \sim \left(- \tilde \rho^2  \dd t^2 + \frac{ L^2 \dd \tilde \rho^2}{\tilde  \rho^2} +  {\tilde \rho^2}\dd \tilde\Omega_3^2 \right)+\\+
 L^2\bigg[\frac{1}{6}\big(\dd \theta^2 + \sin^2  \theta \dd  \phi^2 \big) + \frac 1 6 \big(\dd \hat\theta^2 
+ \sin^2 \hat \theta \dd\hat \phi^2 \big) + \frac 1 9 \big( \dd \hat\psi + \cos \hat \theta \dd \hat \phi +\cos \theta \dd \phi \big)^2\bigg]
\end{multline}
where $\phi$ is the (properly rescaled) polar angle in the $x^1,x^2$ plane. 
We can solve the equations to the next order assuming the behaviour
\begin{gather*}
	m \sim \frac 3 2\frac 1{y^2} + \frac{m_2(\theta)}{y^4}\quad
	n \sim \frac 23\frac{L^4}{y^4}+\frac{n_3(\theta)}{y^6}\quad p \sim \frac 49 \frac{L^4}{y^4}+\frac{p_3(\theta)}{y^6}\\
	T\sim \frac {y}{L^2} \sqrt 6 \sin(\theta) \left(\tan\frac \theta 2\right)^{-1/6}+ \frac{t_1(\theta)}{y}
\end{gather*}
Regularity and coordinate redefinitions now reduce the seven integrations constants to three, $J,Q,B$. The extra constant of integration measures the total flux under the $U(1)$ gauge fields obtained by the reduction of the 5-form $F_{(5)}$ over a non trivial 3-cycle of $T^{1,1}$. This gauge field is associated to the baryon charge in the gauge theory. Baryons can be constructed from operators of the form
\begin{equation}
\epsilon^{\alpha_1\cdots\alpha_N}\epsilon^{\beta_1\cdots\beta_N}\mathcal D_{lm}^{k_1\cdots k_N}
 \prod_{i=1}^N B^{k_i}_{\phantom k \alpha_i \beta_i}
\end{equation}
where  $\alpha_i,\beta_i$ are gauge indexes, $k_i$ are flavour indexes and $D_{lm}^{k_1\cdots k_N}$ is a proper $SU(2)$ Clebsch Gordon coefficient.
\section{Conclusions}
We have presented a parametrisation of exact solutions of Type II B Supergravity dual to (at least) 1/8 BPS chiral primaries in the $\mathcal N=4$ SCFT or to half supersymmetric states in a certain class of $\mathcal N =1$ Quiver gauge theories. These realise to the fully backreacted geometry of giant gravitons and dibaryons in the corresponding gauge theories. The correct matching of charges and satisfaction of the BPS bound is a first non trivial check on the consistency of the derivation. A deeper study of boundary conditions at $y=0$ and the generic $Y^{p,q}$ case is desirable and still work in progress.

\end{document}